\documentclass[aps,twocolumn,showpacs]{revtex4}

\usepackage{graphicx}
\usepackage{amsfonts}
\usepackage{amsmath}

\begin{document}

\title{Repeating 3-body collisions in a trap and the evaluation of
interactions of neutral particles}

\author{C. G. Bao}

\affiliation{Center of Theoretical Nuclear Physics, National
Laboratory of Heavy Ion Collisions, Lanzhou, 73000, P.R. China
 \\
State Key Laboratory of Optoelectronic Materials and Technologies,
Sun Yat-Sen University, Guangzhou, 510275, P.R. China}

\begin{abstract}
A model of a device is proposed and related theoretical calculation
is performed to study the weak interactions among neutral atoms and
molecules. In this model 3-body collisions among the neutral
particles occur repeatedly in a trap. Results of calculation
demonstrate that information on interaction can be obtained by
observing the time-dependent densities of the system.
\end{abstract}

\maketitle

\section{Introduction}

An important way to understand the interactions among particles is
via the study of scattering. In related experiments, the initial
status of a scattering state is required to be precisely controlled.
This is relatively easy for charged incident particles because their
initial momentum can be tuned by adjusting the electromagnetic
forces imposing on them. For neutral incident particles, the initial
momentum is in general difficult to control precisely. However, the
recent progress in the technology of trapping neutral atoms via
optical trap might open a new way for studying the scattering of
neutral particles with precisely controllable initial status
\cite{r_SJ1998,r_BM2001,r_WP2009}. In this paper a model of a device
is proposed and related theoretical calculation is performed to show
how the mentioned scattering is realized. It turns out that in this
device, as we shall see, the collisions among particles occur
regularly and repeatedly. Thereby the effect of each individual
collision can be accumulated. This would be helpful for the
understanding of the very weak interactions among neutral atoms
(molecules).

Traditionally, the scatterings would at most have two incident
channels (say, in the experiments with head-on colliders). However,
in the following device, three or more incident channels can be
realized. As an example, a three-body scattering with three incident
channels is chosen to be studied. This is a generalization of a
previous work on a two-body scattering in a trap with two incident
channels \cite{r_LZB2009}.

It is assumed that, in the beginning, there are three narrow optical
traps located at the three vertexes of a regular triangle, and each
optical trap provides a harmonic confinement. The total potential is
$U_{p}(\mathbf{r})=\frac{1}{2}M\omega
_{p}^{2}\sum_{j=1}^{3}|(\mathbf{r-a}_{j})|^{2}$, where
$\mathbf{a}_{j}$ points from the origin to the $j-th$ vertexes, and
$M$ is the mass of a particle. When the center of the triangle is
placed at the origin, $|\mathbf{a}_{j}|=$ $a$. It is further assumed
that each trap contains an atom in the ground state of a harmonic
oscillator, the three atoms are identical bosons with spin zero, and
$\omega _{p}$ is large enough so that the atoms are well localized
initially. Suddenly the three narrow traps are cancelled. Instead, a
broader new trap located at the origin
$U_{evol}(r)=\frac{1}{2}M\omega ^{2}r^{2}$ is created, $\omega
<\omega _{p}$. Then, the system begins to evolve. The evolution is
affected not only by $U_{evol}(r)$ but also by the atom-atom
interaction $V(|\mathbf{r}_{i}\mathbf{-r}_{j}|)$. In what follows
the details of the evolution is studied, three-body head-on
collisions occurring repeatedly are found, and the effect of
interaction is demonstrated.

\section{Initial state}

We shall use $\hbar \omega $ and $\sqrt{\hbar /M\omega }$ as units
of energy and length. The symmetrized and normalized initial state
\begin{eqnarray}
 \Psi _{I}
 &=&\frac{1}{\sqrt{6}}
  (\frac{\eta }{\pi })^{9/4} \nonumber \\
 &&\times
  \sum_{P}
   e^{
    -\frac{\eta}{2}
    (|\mathbf{r}_{p_{1}}\mathbf{-a}_{1}|^{2}
    +|\mathbf{r}_{p_{2}}\mathbf{-a}_{2}|^{2}
    +|\mathbf{r}_{p_{3}}\mathbf{-a}_{3}|^{2})}
\end{eqnarray}
where $\sum_{P}$ implies a summation over the permutations
$p_{1}p_{2}p_{3}$, and $\eta =\omega _{p}/\omega$. Without loss of
generality, $\mathbf{a}_{3}$ is given lying along the Z-axis, while
the triangle is given lying on the X-Z plane. For convenience, three
sets of Jacobi coordinates denoted by $\alpha$, $\beta$, and
$\gamma$, respectively, are introduced. The coordinates of the
$\alpha $ set are defined as, $\mathbf{r}=\mathbf{r}_{2}-\mathbf{r}
_{1}$,
$\mathbf{R}=\mathbf{r}_{3}-(\mathbf{r}_{1}+\mathbf{r}_{2})/2$, and
$\mathbf{R}_{c}=(\mathbf{r}_{1}+\mathbf{r}_{2}+\mathbf{r}_{3})/3$.
The other two sets can be obtained from the $\alpha$ set by cyclic
permutations. In terms of the $\alpha$ set, we introduce the
harmonic oscillator (h.o.) states $\phi _{nlm}^{(\mu
)}(\mathbf{s})\equiv f_{nl}^{(\mu )}(s)Y_{lm}(\widehat{s})$ as basis
functions, where $\mathbf{s}=\mathbf{r}$, $\mathbf{R}$, or
$\mathbf{R}_{c}$. They are normalized eigenstates of the Hamiltonian
$-\frac{1}{2\mu }\nabla _{\mathbf{S}}^{2}+\frac{1}{2}\mu S^{2}$ with
the eigenenergy $2n+l+3/2$ and with the angular momentum $l$ and its
Z-component $m$, where $\mu =1/2$, $2/3$, and $3$ when
$\mathbf{s}=\mathbf{r}$, $\mathbf{R}$, and $\mathbf{R}_{c}$,
respectively. Then the initial state can be expanded as
\begin{equation}
 \Psi _{I}
 =\frac{1}{\sqrt{6}}
  \sum_{N_{c}}
   c_{N_{c}}
   \phi_{N_{c}00}^{(3)}
   (\mathbf{R}_{c}) \cdot
  \sum_{J,m,\Pi,q}
   G_{Jm\Pi q}
   \Phi_{Jm\Pi q}(\mathbf{r},\mathbf{R})
 \label{e2_PsiI}
\end{equation}
where the first factor is for the c.m. motion which is completely
separated from the internal motion,
\begin{equation}
 c_{N_{c}}
 =\sqrt{4\pi}
  (\frac{\eta}{\pi})^{9/4}
  \int
   R_{c}^{2} dR_{c}
   f_{N_{c}0}^{(3)}(R_{c})
   e^{-\frac{3\eta}{2}R_{c}^{2}}
\end{equation}
The notation $q$ denotes a set of quantum numbers $n,l,\ N$ and $L$,
and $\Pi =(-1)^{l+L}$ is the parity,
\begin{equation}
 \Phi _{Jm\Pi q}(\mathbf{r,R})
 \equiv [ \phi _{nl}^{(1/2)}(\mathbf{r})
          \phi _{NL}^{(2/3)}(\mathbf{R})_{Jm} ]
 \label{e4_PsiJmPq}
\end{equation}
where $l$ and $L$ are coupled to $J$ and $m$.
\begin{eqnarray}
 G_{Jm\Pi q}
 &=&\sum_{q'}
   a_{n'l'm}
   b_{N'L'}
   C_{l'm,L'0}^{Jm}
   (1+(-1)^{l'}) \nonumber \\
  &&\times [ \delta_{qq'}
    +\mathcal{A}_{q}^{q'J}(\beta \rightarrow \alpha )
    +\mathcal{A}_{q}^{q'J}(\gamma \rightarrow \alpha ) ]
\end{eqnarray}
where
\begin{eqnarray}
 a_{nlm}
 &=&\int
   r^{2} dr\
   f_{nl}^{(1/2)}(r)\
   e^{-\frac{\eta }{4}r^{2}} \nonumber \\
 &&\times \int d\widehat{r}\
   Y_{lm}^{\ast }(\widehat{r})
   e^{\frac{\eta }{2}(\sqrt{3} \arcsin\theta_{r} \cos\phi_{r} - 3a^{2}/2)}
\end{eqnarray}
\begin{eqnarray}
 b_{NL}
 &=&\int R^{2}dR\
   f_{NL}^{(2/3)}(R)\
   e^{-\frac{\eta }{3}R^{2}} \nonumber \\
 &&\times \int d\widehat{R}\
  Y_{L0}^{\ast }(\widehat{R})
  e^{\eta(\arccos\theta_{R} - 3a^{2}/4)}
\end{eqnarray}
$C_{l'm,L'0}^{Jm}$ is the Clebsch-Gordan coefficients, $q'$ is for
the set $n'$, $l'$, $N'$ and $L'$ to be summed up, $\theta _{r}$ and
$\phi _{r}$ are the spherical polar coordinates of $\mathbf{r}$, and
so on.
\begin{widetext}
\begin{equation}
 \mathcal{A}_{q}^{q'J}(\beta \rightarrow \alpha )
 \equiv \langle \ [
  \phi _{nl}^{(1/2)}(\mathbf{r)}
  \phi _{NL}^{(2/3)}(\mathbf{R)}]_{J}
 \ | \ [
  \phi _{n'l'}^{(1/2)}(\mathbf{r}^{\beta })
  \phi _{N'L'}^{(2/3)}(\mathbf{R}^{\beta })]_{J} \ \rangle
\end{equation}
\end{widetext}
is the bracket of transformation between the $\beta -$ and $\alpha
-$sets (the superscript $\alpha $ is usually ignored), which is
called the Talmi-Moshinsky (T-M) coefficients. Their analytical
expression can be found in \cite{r_TW1981,r_BM1966,r_BTA1960}. In
general, the symmetrization would cause the appearance of all three
set of coordinates. However, by using the T-M coefficients,
Eq.~(\ref{e2_PsiI}) contains only the $\alpha -$set so as to
facilitate greatly the calculation. Incidentally, since $\Psi _{I}$\
is symmetrized, $l$ (included in $q$) of Eq.~(\ref{e2_PsiI}) must be
even and therefore $(-1)^{L}=\Pi$. In principle, the right side of
Eq.~(\ref{e2_PsiI}) should contain infinite terms. However, the
overlap between $\Psi _{I}$ and higher h.o. states are very small.
Say, if, $n$, $N$, $N_{c}$ are all smaller than 12 and $l$, $L$,
$L_{c}$ are all smaller than $23$, with the parameters specified
below, the overlap of the right side of Eq.~(\ref{e2_PsiI}) with
itself is equal to $0.99997$. This implies that higher h.o. states
can be safely ignored.

\section{Hamiltonian and its eigenstates}

The evolution is governed by the Hamiltonian $H_{evol}$ containing
$U_{evol}(r)$\ and the interaction. When the Jacobi coordinates are
used it can be separated as
\begin{equation}
 H_{evol}=H_{c}+H_{in}
\end{equation}
where $H_{c}=-\frac{1}{6}\nabla
_{\mathbf{R}_{c}}^{2}+\frac{3}{2}R_{c}^{2}$ describes the c.m.
motion, and
\begin{equation}
 H_{in}
 =-\nabla _{\mathbf{r}}^{2}
  +\frac{1}{4}r^{2}
  -\frac{3}{4}\nabla _{\mathbf{R}}^{2}
  +\frac{1}{3}R^{2}
  +\sum_{i<j}V(|\mathbf{r}_{i}-\mathbf{r}_{j}|)
\end{equation}
describes the internal motion. Since the eigenstates of $H_{c}$ are
well known, if the eigenstates of $H_{in}$ are also know, the
evolution starting from any initial state can be understood. In what
follows, the symmetrized eigenstates are obtained via a
diagonalization of $H_{in}$ in a limited space. In this way, only
approximate solutions can be obtained. Then, we increase the
dimension of the space until a better convergency is achieved.

When $\Phi _{Jm\Pi q}(\mathbf{r,R})$ as defined in
Eq.~(\ref{e4_PsiJmPq}) are used as basis functions (where $l$ is
restricted to be even as mentioned), the matrix elements of
$H_{in}$\ is
\begin{widetext}
\begin{eqnarray}
 &&\langle \ \Phi_{J'm'\Pi'q'}(\mathbf{r},\mathbf{R})
 | \ H_{in} \ | \
 \Phi _{Jm\Pi q}(\mathbf{r},\mathbf{R}) \ \rangle \nonumber \\
 &=&\delta _{J'J}
  \delta _{m'm}
  \delta _{\Pi'\Pi }\
  [\ \delta_{q'q} (2n+2N+l+L+3)
   +\delta_{l'l}
    \delta_{N'N}
    \delta_{L'L} V_{n'nl} \nonumber \\
   &&+2\sum_{q'',q'''}
      \delta_{l'''l''}
      \delta_{N'''N''}
      \delta_{L'''L''}
      \mathcal{A}_{q'''}^{q'J}(\alpha \rightarrow \beta )
      \mathcal{A}_{q''}^{qJ}(\alpha \rightarrow \beta )\
      V_{n'''n''l''}\ ]
\end{eqnarray}
\end{widetext}
where $q''$\ denotes the set $(n''l''N''L'')$, and the
implication of $q'''$ is alike,
\begin{equation}
 V_{n'nl}=\int r^{2}dr\ f_{n'l}^{(1/2)}(r)\ V(r)f_{nl}^{(1/2)}(r)
\end{equation}
To control the size of the space, a number $N_{0}$ is introduced and
$2n+2N+l+L\leq N_{0}$ is required for all the basis functions. There
are two choices to obtain symmetrized eigenstates of $H_{in}$. In
the first choice, the set $\Phi_{Jm\Pi q}(\mathbf{r},\mathbf{R})$ is
firstly symmetrized and orthonormalized before carrying on the
diagonalization. However, this procedure is complicated. Therefore
we make the second choice, in which the set $\Phi _{Jm\Pi
q}(\mathbf{r},\mathbf{R})$ is simply used without symmetrization but
with the requirement that $l$ must be even and all the basis
functions satisfying $2n+2N+l+L\leq N_{0}$ are included without
missing. This requirement assures that the space is close under
permutations, and no basis functions that will contribute to the
symmetrized eigenstates would be missed, unless they are too high to
have $2n+2N+l+L>N_{0}$. However, in this choice, a number of
unphysical eigenstates with confused symmetry will emerge together
with those with correct symmetry. Therefore a discrimination is
needed as shown below.

Let an eigenstate be denoted as $\Psi _{Jm\Pi i}$ where $i$ is a
serial number of the $Jm\Pi -$series. Expanding in terms of the
basis functions,
\begin{equation}
 \Psi _{Jm\Pi i}
 =\sum_{q}
   \mathcal{B}_{iq}^{Jm\Pi }
   \Phi _{Jm\Pi q}(\mathbf{r},\mathbf{R })
\end{equation}
where the $l$ in $q$ must be even, and the coefficients
$\mathcal{B}_{iq}^{Jm\Pi }$ can be directly known from the
diagonalization of $H_{in}$. If $\Psi _{Jm\Pi i}$ is correctly
symmetrized, the coefficients would obey
\begin{eqnarray}
 &&\sum_{q}
  \mathcal{B}_{iq}^{Jm\Pi }
  \mathcal{A}_{q'}^{qJ}(\beta \rightarrow \alpha )
 =\mathcal{B}_{iq'}^{Jm\Pi },\nonumber \\
 \ \ \ \ \ \ && \mbox{(for all the $q'$ with $l'$ even)}
 \label{e14_BAeven}
\end{eqnarray}
and
\begin{equation}
 \sum_{q}
  \mathcal{B}_{iq}^{Jm\Pi }
  \mathcal{A}_{q'}^{qJ}(\beta \rightarrow \alpha )
 =0,\ \ \
  \mbox{(for all the $q'$ with $l'$ odd)}
 \label{e15_BAodd}
\end{equation}
In the summations of Eqs.~(\ref{e14_BAeven}) and (\ref{e15_BAodd}),
$l$ (in $q$) is restricted to be even. With the help of
Eqs.~(\ref{e14_BAeven}) and (\ref{e15_BAodd}), the states with
confused symmetry can be discriminated and dropped, and all the
symmetrized eigenstates under the restriction caused by $N_{0}$ can
be extracted without missing. They are one-to-one identical to those
obtained via the first choice if the same $N_{0}$ are used. In what
follows $\Psi _{Jm\Pi i}$ denotes only the symmetrized eigenstate,
and the associated energy is denoted by $E_{J\Pi i}$.

\section{Evolution and the repeating 3-body collisions}

With the eigenstates it is straight forward to obtain the
time-dependent solution of $H_{evol}$\ as
\begin{equation}
 \Psi (t)
 =e^{-iH_{evol}\ \tau } \Psi _{I}
 \equiv \Psi _{c}(\mathbf{R}_{c},t)
  \Psi _{in}(\mathbf{r},\mathbf{R},t)
 \label{e16_Psit}
\end{equation}
\begin{equation}
 \Psi _{c}(\mathbf{R}_{c},t)
 =\frac{1}{\sqrt{6}}
  \sum_{N_{c}}
   c_{N_{c}}
   e^{-i\tau (2N_{c}+3/2)\ }
   \phi _{N_{c}00}^{(3)}(\mathbf{R}_{c})
\end{equation}
\begin{eqnarray}
 \Psi _{in}(\mathbf{r},\mathbf{R},t)
 &=&\sum_{J,m,\Pi,q}
   G_{Jm\Pi q}
   \sum_{i}
    e^{-i\tau E_{J\Pi i}\ } \nonumber \\
    &&\times | \
     \Psi _{Jm\Pi i} \ \rangle \langle \
     \Psi _{Jm\Pi i} \ | \ \Phi _{Jm\Pi q}\rangle \nonumber \\
 &=&\sum_{J,m,\Pi,q'}
   \mathcal{D}_{q'}^{Jm\Pi }(t)\
   \Phi _{Jm\Pi q'}(\mathbf{r},\mathbf{R})
 \label{e18_PsiinrRt}
\end{eqnarray}
where
\begin{equation}
 \mathcal{D}_{q'}^{Jm\Pi }(t)
 =\sum_{i,q}
   G_{Jm\Pi q}
   e^{-i\tau E_{J\Pi i}}\
   \mathcal{B}_{iq}^{Jm\Pi }\mathcal{B}_{iq'}^{Jm\Pi }
\end{equation}
and $\tau =\omega t$. Obviously, Eqs.~(\ref{e16_Psit}) to
(\ref{e18_PsiinrRt}) give only an approximate solution because the
set $\Psi _{Jm\Pi i}$ obtained via diagonalization each would
deviate more or less from the corresponding exact eigenstate, and
because only finite number of $\Psi _{Jm\Pi i}$ are used in the
expansion. However, it is believed that, when the number of basis
functions becomes larger and larger, the above $\Psi (t)$ would be
closer and closer to the exact solution. The crucial point is the
convergency. In this paper the interaction is assumed to be weak. It
turns out that, in this case, the convergency is satisfying as shown
below.

We shall demonstrate that the evolution is a repeating 3-body collisions,
and the effect of interaction will be also shown. For these purposes, we
extract the following quantities from $\Psi (t)$.

(i) The density $\rho _{r}(r,t)\equiv \int d\Omega _{r}\Psi ^{\ast }(t)\Psi
(t),$ where the integration covers all the degrees of freedom except $dr$.
Therefore $\int dr\ \rho _{r}(r,t)=1.$ Obviously, $\rho _{r}(r,t)$ is the
probability density that the inter-distance is $r$.

(ii) The density $\rho _{R}(R,t)$ fulfilling $\int dR\ \rho _{R}(R,t)=1$

(iii) The density $\rho _{\theta }(\theta ,t)$ fulfilling $\int \sin
\theta \cdot d\theta \ \rho _{\theta }(\theta ,t)=1$, where $\theta
$ is the angle between $\mathbf{r}$ and $\mathbf{R}$.

(iv) The density $\zeta(\mathbf{R},t)$ fulfilling \bigskip $\int d
\mathbf{R}\ \zeta(\mathbf{R},t)=1$.

It was found that the c.m. is distributed very close to the origin,
therefore $\mathbf{R}$ is approximately proportional to
$\mathbf{r}_{3}$. Therefore, the behavior of the particle 3 can be
roughly understood via $\zeta(\mathbf{R},t)$. Incidentally, the
behaviors of all the particles are the same due to the
symmetrization. The analytical expression of these densities are
given in the appendix.

\begin{figure}[htbp]
 \centering \resizebox{0.8\columnwidth}{!}{
 \includegraphics{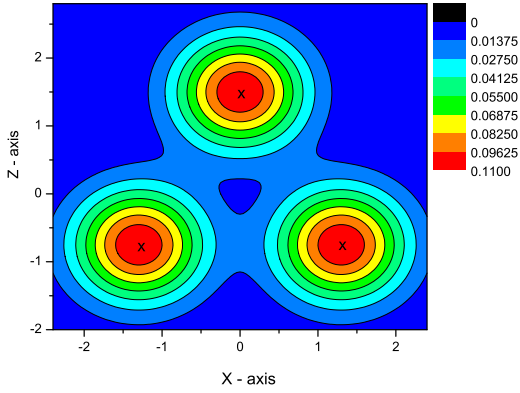} }
\caption{Contour diagram of the one-body density
$\rho_{1}(\mathbf{r}_{1})$ of the initial state plotted in the X-Z
plane. $\eta =1.5$ and $|\mathbf{a}|=1.5$ are adopted. Each maximum
is marked with a $\times $. The three particles form a regular
triangle with side-length $\sim 2.6$ initially. (Color online)}
\end{figure}

In order to have numerical results, as an example, the interaction
is assumed to be a repulsive core as $V(r)=V_{0}$ if $r\leq 0.4$, or
zero if $r>0.4$, where $V_{0}$ is a constant to be given. The other
parameters are chosen as $\eta =1.5$, $|\mathbf{a}|=1.5$, and
$N_{0}=20$. To show the initial localization of the particles, we
define the one-body density of the initial state as $\rho
_{1}(\mathbf{r}_{1})\equiv \int d\mathbf{r}_{2}d \mathbf{r}_{3}\Psi
_{I}^{\ast }\Psi _{I}$. This density is plotted in Fig.1. Starting
from $\Psi _{I}$, the details of evolution are given as follows.

\begin{figure}[htbp]
 \centering \resizebox{0.8\columnwidth}{!}{
 \includegraphics{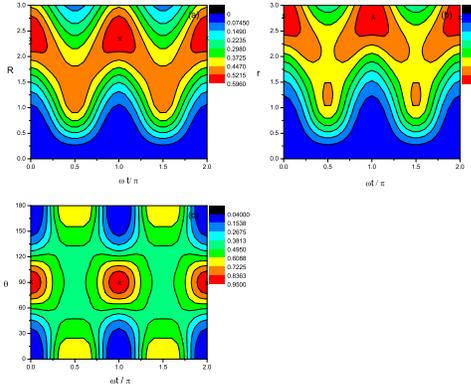} }
\caption{Contour plots of $\rho _{R}(R,t)$, $\rho _{r}(r,t)$, and
$\rho _{\theta }(\theta ,t)$ given in (a), (b), and (c),
respectively, with $\eta =1.5$, $|\mathbf{a}|=1.5$, $V_{0}=0.08$ and
$N_{0}=20$. Each maximum is marked with a $\times $. (Color online)}
\end{figure}

$\rho _{R}(R,t)$, $\rho _{r}(r,t)$, and $\rho _{\theta }(\theta ,t)$
against $t$\ in the earliest stage of evolution is plotted in Fig2a
to 2c. Where $V_{0}=0.08$ is assumed, and $\omega t$ is from $0$ to
$2\pi $ (say, if $\omega =1000\times 2\pi$, then $t$ is from $0$ to
$0.001\sec $). When $t=0$, the inter-distances among the particles
are about $2.6$ as shown in Fig.1 and 2b. When the evolution begins,
the peaks of $\rho _{r}$ and $\rho _{R}$ move inward synchronously,
while $\rho _{\theta }$ remains to peak at $90^{\circ }$. It implies
a contraction of the regular triangle. Accordingly, the three
particles rush towards the center leading to a 3-body head-on
collision. When $\omega t$ is close to $\pi /2$, the inter-distances
become much shorter, and $\rho _{\theta }$ becomes nearly uniform.
It implies that the geometric character (i.e., the regular triangle)
will be spoiled when the particles are close to each other.

\begin{figure}[htbp]
 \centering \resizebox{0.8\columnwidth}{!}{
 \includegraphics{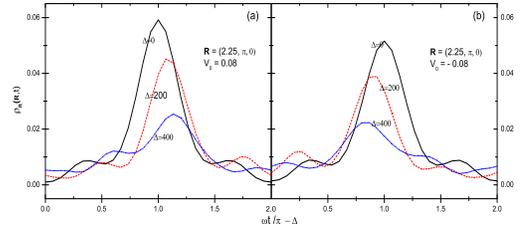} }
\caption{$\zeta (\mathbf{R},t)$ against $t$ given in three selected
intervals of $t$. $\mathbf{R}$ is fixed at $(2.25,\pi ,0)$ (i.e.,
$\mathbf{R}$ is lying along the negative Z-axis with $R=2.25$.
Meanwhile particle 3 is close to the opposite end of the up-vertex
of the initial triangle). The three curves are associated with the
three choices of $\Delta $, namely, $0$, $200,$ and $400$. When
$\Delta =0$, the interval of $\omega t$ is [$0,2\pi $] (solid
curve). When $\Delta =200$, it is $[200\pi,\ 202\pi]$ (dash curve).
When $\Delta =400$, it is $[400\pi,\ 402\pi]$ (dash-dot-dot curve).
$V_{0}=0.08$ (a) and $-0.08$ (b). The other parameters are the same
as in Fig.2. (Color online)}
\end{figure}

After the first collision, the particles move outward. When $\omega
t=\pi$, each particle will be close to the opposite end of its
initial position as shown by the solid curve of Fig.3a. Where
$\zeta(\mathbf{R},t)$ is given at the point $\mathbf{R}\equiv
(R,\theta _{R},\phi _{R})=(2.25,\pi ,0)$ and $\omega t$ is given in
the interval $[0,2\pi ]$. The sharp peak of the solid curve
demonstrates clearly that a particle arrives at the opposite end
when $\omega t\sim \pi$. Due to the symmetrization of the wave
function, this is also true for other two particles. Then, the above
process begins to reverse. When $\omega t=2\pi$, the system recovers
its initial status nearly. If the interaction is removed, the
recovery is complete and the system will undergo an exact periodic
motion with the period $2\pi /\omega$. The motion is characterized
by the repeating head-on 3-body collisions. each occurs once within
the interval $\pi /\omega$.

However, due to the interaction, the recovery is not exact. When the
time goes on the effect of the weak interaction will accumulate and
gradually emerge. This is shown by the dash-curve of Fig.3a which
describes the behavior of $\zeta (\mathbf{R},t)$ after $200$ rounds
of head-on collisions. Where the peak is lower than that of the
solid curve. It implies that the density is diffusing. Furthermore,
the peak of the dash-curve has shifted a little right implying that
the arrival is a little delayed. The diffusion and the delay will
become more explicit when $t$ is larger as shown by the dash-dot-dot
curve. When the interaction is attractive, the above repeating
3-body collisions remain, and the densities remain to be diffusing.
However, instead of a delay, the peak will arrive at the end earlier
as shown in Fig.3b, where $V_{0}=-0.08$. Due to the diffusion of the
density as shown in Fig.3, the phenomenon of repeating 3-body
collisions will become ambiguous when $t$ is sufficiently large.

\begin{figure}[htbp]
 \centering \resizebox{0.8\columnwidth}{!}{
 \includegraphics{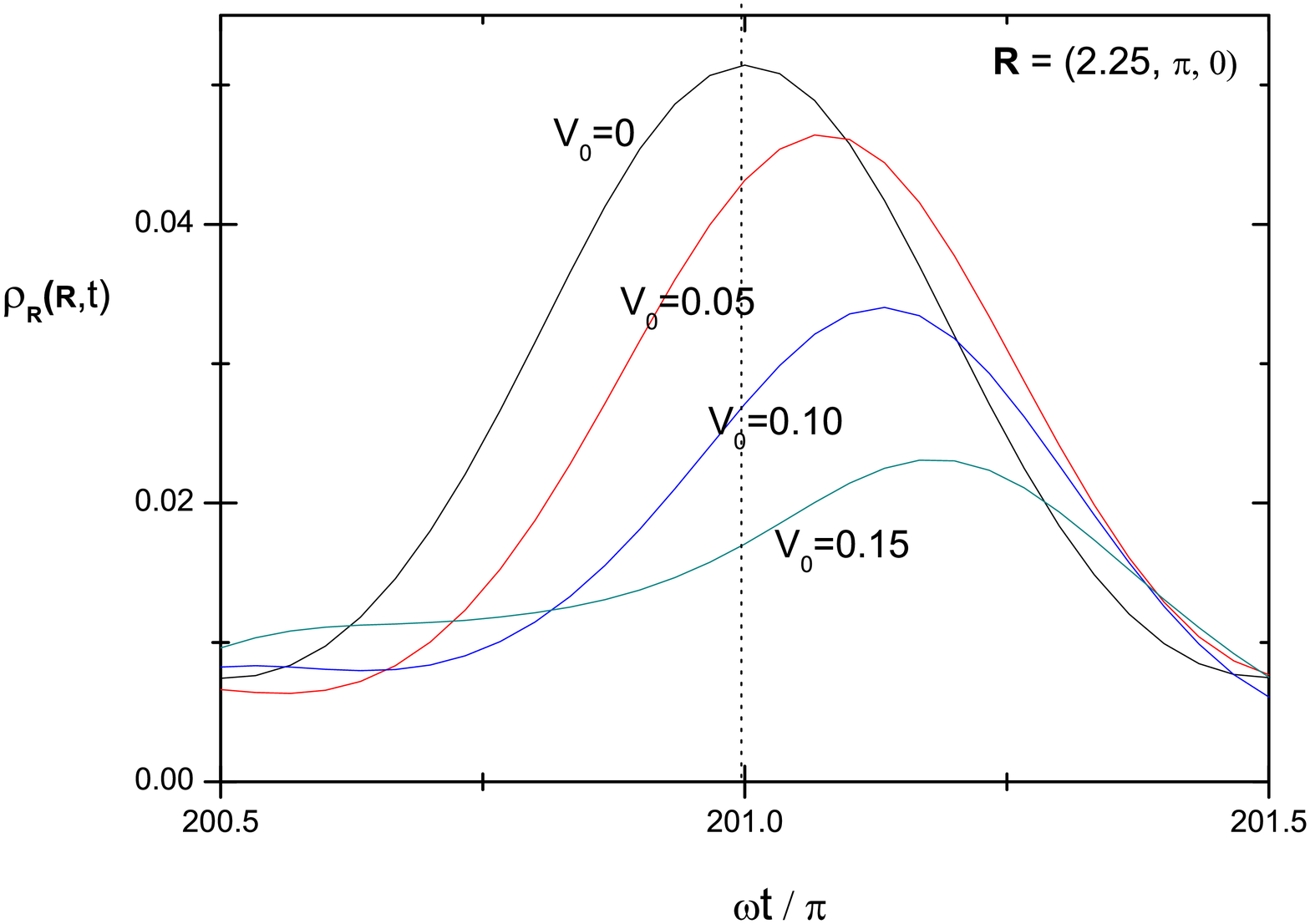} }
\caption{$\zeta (\mathbf{R},t)$ against $t$\ in an interval close to
201$\pi /\omega $. $\mathbf{R}$ is fixed at $(2.25,\pi ,0)$. $V_{0}$
is given at four values marked by the curves. The other parameters
are the same as in Fig.2. (Color online)}
\end{figure}

When $V_{0}$ is given at a number of values and $\mathbf{R}$ is
fixed at $(2.25,\pi ,0)$, $\zeta(\mathbf{R},t)$ against $\omega t$
is plotted in Fig.4, in which $\omega t$ varies in an interval close
to $201\pi$. In this figure both the locations and the heights of
the peaks depend on the strength $V_{0}$. In general, a larger
$|V_{0}|$ will cause a stronger diffusion and therefore a shorter
life of the phenomenon of clear repeating 3-body collisions, and a
more positive (negative) $V_{0}$\ will cause a larger shift of the
peak to the right (left). In addition to $V_{0}$, the evolution
depends also on the range of interaction. Therefore, by observing
the time-dependent densities, information on the interaction can be
obtained.

The accuracy of the above numerical results depends on $N_{0}$, the
number controlling the total number of basis functions. As an
example the values of $\zeta (\mathbf{R},t)$ are listed in Table 1
to show the dependence.
\begin{table}[htbp]
 \caption{The values of $\zeta(\mathbf{R},t)$ with
$\mathbf{R}=(2.25,\pi ,0)$ and with three choices of $N_{0}$. The
other parameters are the same as those for Fig.2}
 \label{tab:1}
 \begin{tabular}{llll}
  \hline
  $N_{0}$ & 16 & 20 & 24 \\
  \hline
  $\zeta (\mathbf{R},31\pi /\omega )$  & 0.0590 & 0.0586 & 0.0586 \\
  $\zeta (\mathbf{R},91\pi /\omega )$  & 0.0547 & 0.0542 & 0.0543 \\
  $\zeta (\mathbf{R},151\pi /\omega )$ & 0.0472 & 0.0468 & 0.0469 \\
  \hline
 \end{tabular}
\end{table}

The convergency shown in this table is satisfying. Thus we conclude that the
numerical results obtained by using $N_{0}=20$ are accurate enough in
qualitative sense.

\section{Summary}

A model of a device containing repeating 3-body collisions is
proposed to study the weak interactions among neutral atoms and
molecules. The advantage is twofold. (i) The initial status can be
precisely controlled. (ii) The weak effect of the interaction can be
accumulated and therefore easier to be detected. Numerical results
support that information on interaction can be thereby extracted.

A crucial point in the device is the initial localization of the particles.
If they are better localized (by increasing $\omega _{p}$) and/or they are
more separated from each other initially (by increasing $|\mathbf{a}_{j}|$),
the repeating 3-body collisions would become more explicit and be maintained
longer. Furthermore, if $H_{evol}$ has a larger $\omega $, the interval
between two successive collisions ($\sim \pi /\omega $) becomes shorter,
therefore the evolution would proceed swifter.

Although the three particles are assumed to be identical bosons, a
generalization to fermions and/or distinguishable particles is straight
forward. In principle, the above device could also be used to study the
three-body forces if they exist.

\bigskip

\begin{acknowledgements}
The support from the NSFC under the grant 10874249 is appreciated.
\end{acknowledgements}

\bigskip

\appendix

\section{Analytical expressions of the time-dependent densities}

The analytical expressions of the time-dependent densities are as follows.
\begin{eqnarray}
 \mbox{(i)} \ \rho_{R}(R,t)
 &=&\frac{1}{6}
  \sum_{N_{c}} (c_{N_{c}})^{2}
  \sum_{J,m,\Pi,q',q}
   \delta_{n'n}
   \delta_{l'l}
   \delta_{L'L} \nonumber \\
 &&\times
   [\mathcal{D}_{q'}^{Jm\Pi }(t)]^{\ast }\
   \mathcal{D}_{q}^{Jm\Pi }(t) \nonumber \\
 &&\times
   R^{2}
   f_{N'L}^{(2/3)}(R)
   f_{NL}^{(2/3)}(R)
\end{eqnarray}
\begin{eqnarray}
 \mbox{(ii)} \ \rho _{r}(r,t)
 &=&\frac{1}{6}
  \sum_{N_{c}} (c_{N_{c}})^{2}
  \sum_{J,m,\Pi,q',q}
   \delta_{N'N}
   \delta_{l'l}
   \delta_{L'L} \nonumber \\
 &&\times
   [\mathcal{D}_{q'}^{Jm\Pi }(t)]^{\ast }\
   \mathcal{D}_{q}^{Jm\Pi }(t) \nonumber \\
 &&\times
   r^{2}
   f_{n'l}^{(1/2)}(r)
   f_{nl}^{(1/2)}(r)
\end{eqnarray}
(iii) For the derivation of $\rho _{\theta }(\theta ,t)$, a
transformation is made so that $\mathbf{r}$ and $\mathbf{R}$ are
transformed to $r,\ R,\ \theta ,\ $\ and the three Euler angles
specifying the orientation of the triangle formed by the three
particles. Then, integrating all the degrees of freedom except
$\theta $, we have
\begin{widetext}
\begin{eqnarray}
 \rho _{\theta }(\theta,t)
 &=&\frac{1}{6}
  \sum_{N_{c}} (c_{N_{c}})^{2}
  \sum_{J,m,\Pi',\Pi,q',q}
   [\mathcal{D}_{q'}^{Jm\Pi'}(t)]^{\ast }\
   \mathcal{D}_{q}^{Jm\Pi }(t)\
   (-1)^{l+l'-J} \nonumber \\
 &&\times
   [(2L^{\prime }+1)(2L+1)(2l^{\prime }+1)(2l+1)]^{1/2} \
   [\int r^{2}dr\ f_{n'l'}^{(1/2)}(r)\ f_{nl}^{(1/2)}(r)] \nonumber \\
 &&\times
   [\int R^{2}dR\ f_{N'L'}^{(2/3)}(R)\ f_{NL}^{(2/3)}(R)] \nonumber \\
 &&\times
   \sum_{\lambda }
    W(ll'LL';\lambda J)
    \sqrt{\frac{\pi}{2\lambda+1}}
    C_{L0,L'0}^{\lambda 0}
    C_{l0,l'0}^{\lambda 0}
    Y_{\lambda 0}(\theta ,0)
\end{eqnarray}
\end{widetext}
where the Wigner and Clebsch-Gordan coefficients are introduced.
\begin{widetext}
\begin{eqnarray}
 \mbox{(iv)} \ \zeta \mathbf{(R},t)
 &=&\frac{1}{6}
  \sum_{N_{c}} (c_{N_{c}})^{2}
  \sum_{J',J,m',m,\Pi',\Pi ,q',q}
   \delta_{n'n}
   \delta_{l'l}
   [\mathcal{D}_{q'}^{J'm'\Pi'}(t)]^{\ast }\
   \mathcal{D}_{q}^{Jm\Pi }(t) \nonumber \\
 &&\times
  \sum_{k}
   [ C_{l'k,L'm'-k}^{J'm'}
     C_{lk,Lm-k}^{Jm}
     Y_{L',m'-k}^{\ast}(\widehat{R})
     Y_{L,m-k}(\widehat{R})] \nonumber \\
 &&\times
     f_{N'L'}^{(2/3)}(R)\
     f_{NL}^{(2/3)}(R)
\end{eqnarray}
\end{widetext}

\end{document}